\documentclass[conference]{IEEEtran}
\IEEEoverridecommandlockouts
\usepackage{cite}
\usepackage{amsmath,amssymb,amsfonts}
\usepackage{wasysym}
\usepackage{algorithm, algorithmic}
\usepackage{graphicx}
\usepackage{textcomp}
\usepackage{subfigure}
\usepackage{listings}

\def\BibTeX{{\rm B\kern-.05em{\sc i\kern-.025em b}\kern-.08em
    T\kern-.1667em\lower.7ex\hbox{E}\kern-.125emX}}
    
\newcommand{\sfunction}[1]{\textsf{\textsc{#1}}}   

\makeatletter
\newcommand*\titleheader[1]{\gdef\@titleheader{#1}}
\AtBeginDocument{%
  \let\st@red@title\@title
  \def\@title{%
    \bgroup\normalfont\large\centering\@titleheader\par\egroup
    \vskip0.1em\st@red@title}
}
\makeatother
\title{WAE: Workload Automation Engine for CDN-specialized Container Orchestration}

\titleheader{2018 IEEE Second International Balkan Conference on Communications and Networking}
\author{maxmin}

\begin{document}

\author{\IEEEauthorblockN{Elif AK\IEEEauthorrefmark{1},
Taner OZDAS\IEEEauthorrefmark{2}, Serkan SEVIM\IEEEauthorrefmark{2} and
Berk CANBERK\IEEEauthorrefmark{1}}
\IEEEauthorblockA{\IEEEauthorrefmark{1} Department of Computer Engineering, Istanbul Technical University, Turkey \\
Email: \{akeli,canberk\}@itu.edu.tr}
\IEEEauthorblockA{\IEEEauthorrefmark{2} MEDIANOVA, Turkey\\
Email: \{taner.ozdas, serkan.sevim\}@medianova.com\\
}}

\maketitle
\begin{abstract}
Content Delivery Network (CDN) has been emerged as a compelling technology to provide efficient and scalable web services even under high client request. However, this leads to a dilemma between minimum deployment cost and robust service under heavy loads. To solve this problem, we propose the Workload Automation Engine (WAE) which enables dynamic resource management, automated scaling and rapid service deployment with least cost for CDN providers. Our modular design uses an algorithm to calculate the optimal assignment of virtual CDN functions such as streaming, progressive delivering and load balancer. In particular, we study on real CDN data which belongs to Medianova CDN Company in Turkey. Also we use \textit{Docker} containerization as an underlying system. The results reveal that our containerized design reduces the latency and deployment cost by 45\% and by 66\%, respectively. Moreover, we obtain roughly 20\% more CPU efficiency and 35\% more utilized network.     

\end{abstract}


\section{Introduction}
Delivering massive volume of content in admissible speed has been a challenge and Content Delivery Network (CDN) try to overcome distributing content to end users with improving the Quality of Service (QoS) \cite{b12}. According to report of Technavio, CDN market is expected to grow at CAGR of 30\% from 2017 to 2021 \cite{b1}. However, in the nature of CDN, it builds huge number of Point of Presences (PoPs) which are distributed geographically \cite{b11} to provide replica content across a wide area. This brings high physical overloading and non-flexible structure to CDN providers, also it has become difficult to compensate and manage this increasing CDN market demand.

\begin{figure}
    \centering
    \subfigure[Client requests log for small edge server in 1-day period]
    {
        \includegraphics[width=\linewidth]{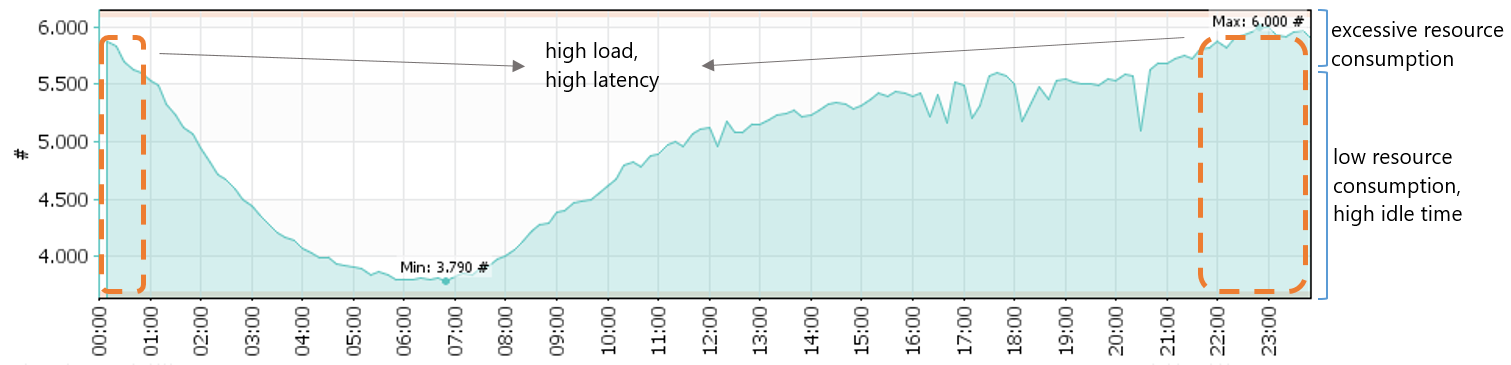}
        \label{fig:image_connection}
    }
    \subfigure[Client requests log for large edge server in 1-day period]
    {
        \includegraphics[width=1.01\linewidth]{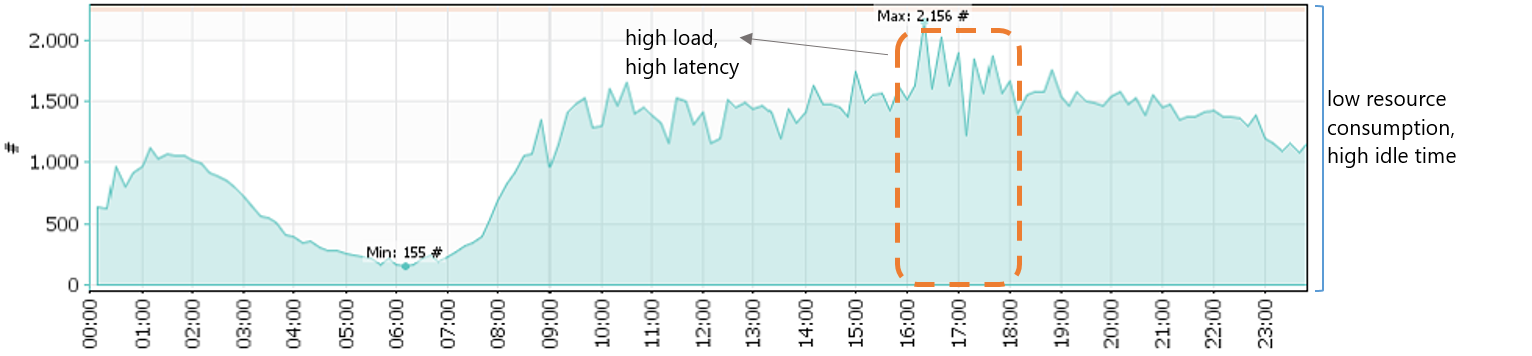}
        \label{fig:large_connection}
    }
    \subfigure[Client requests log for stream edge server in 1-day period]
    {
        \includegraphics[width=\linewidth]{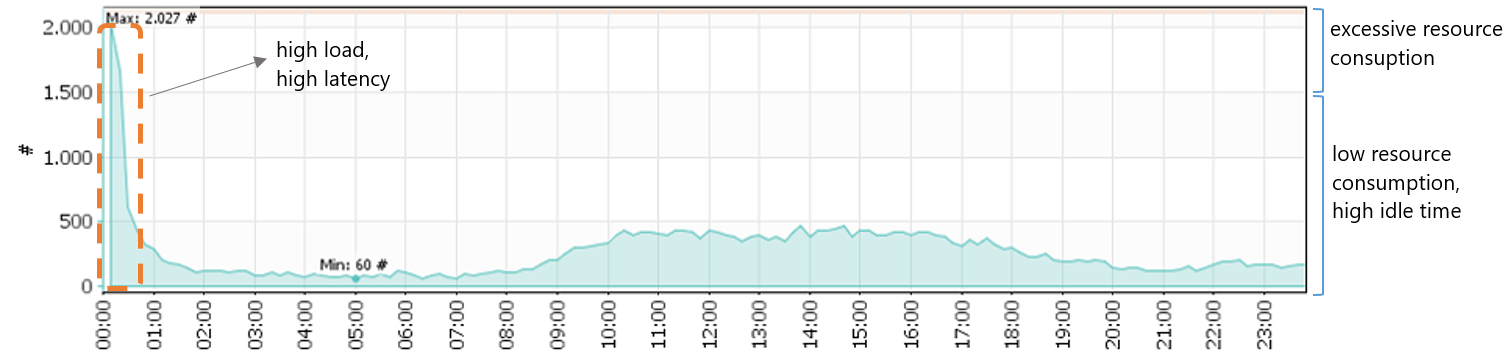}
        \label{fig:stream_connection}
    }
    \caption{Active connection count with respect to edge servers }
    \label{fig:connection_comparison}
\end{figure}

More specifically, there are three key issues in CDN to be addressed: (i) stable and low latency even if under high load, (ii) balanced resource management that decreases idle time of machines, and (iii) minimizing deployment cost without effecting latency. Moreover, Fig. \ref{fig:connection_comparison} shows our observations on number of HTTP connections for 1-day period. This figure is direct network trace taken from Vienna branch of Medianova, one of the CDN companies in Turkey. As seen in Fig. \ref{fig:connection_comparison}, the intervals in which high load cause excessive latency vary by day of hours. For example, in Fig. \ref{fig:stream_connection}, the number of users who requests the live streaming is in peak demand at the beginning of the day (00.00). However, in the same interval, users who demand the large files are relatively less as seen in Fig. \ref{fig:large_connection}.  Such heterogeneous client requests that rapidly increase the latency indicate the importance of dynamically changeable CDN functionalities to compensate high load.     

On the other hand, none of CDN functionalities work concurrently under high resource requirements. So it leads to redundant idle time and non-effective resource usage on the machines. For example, in Fig. \ref{fig:connection_comparison}, the client requests which lead to excessive resource consumption make up almost 4 percentage of the day. In other words, hardware based CDN functionalities remain idle in most of the day.       

With these perspectives, current literature focuses on the network function virtualization (NFV) to put aside aforementioned hardware restrictions by giving chance to software networking. In \cite{b2}, the authors study on optimal placement of virtual CDN (vCDN) components to provide high streaming quality. Herbaut et al. \cite{b3} focus on the virtualization for ISP-CDN collaboration. Frangoudis et. al. \cite{b4} proposes a vCDN model over network operators. Also, \cite{b4} and \cite{b9} provide detail comparison on Docker containers with KVM hypervisor in terms of startup time, request throughput and response. The results show that unlike the traditional virtual machines (VMs), lightweight containers give better results in vCDN service. Moreover, there are many studies on container orchestration to provide service on demand. Tosatto et. al. \cite{b5} assess the challenges in container orchestration. In \cite{b6}, genetic algorithm is proposed for optimized container allocation. In \cite{b7}, linear programing based solution is proposed for containerized applications. Also, \cite{b8} provide detail comparison on container orchestration tools in industry. \cite{b10} proposes an orchestration model to improve insufficiency of \textit{Kubernetes}, one of the orchestration tools in industry. These are the specific and comprehensive works which motivate our own work. In most of these recent studies, CDN-specialized container orchestration has not been covered in terms of resource provisioning. However, orchestration solution is necessary to provide steady service for end-user and optimized resource usage for CDN providers. 

Consequently, keeping these studies in mind, we propose a container based virtualized architecture which provides resilient resource provisioning for CDN network. Our proposed design consists of Docker containerization as the infrastructural layout. Using the heterogeneous client requests and resource capabilities as inputs, the orchestration module re-assign CDN roles dynamically to meet client request with lower latency. The architecture includes helper machines communicating with the WAE through interfaces to supply such inputs to the orchestration module. The orchestration module periodically calculates new arrangements. Also the proposed model enables CDN providers to build own PoPs with least physical machines, i.e., with least cost. Because our design aims to use underlying hardware with full performance to minimize idle time of the machines. To the best of our knowledge, this is the first study on the CDN-specialized container orchestration framework with resource provisioning. The main contributions of this paper includes the following:
\begin{itemize}
\item Unlike the traditional CDN architecture, we design and implement virtualized CDN network which enable PoPs to assign own network functions dynamically. By this way, we obtain lightweight system, efficient resource usage, reduced deployment time and auto-manageable PoPs.
\item Our implementation has all network functions in CDN also includes WAE that responsible for container (VNF) assignments and resource provisioning on the contrary other virtualized CDN designs. 
\item WAE specializes container orchestration to the CDN thanks to the included five modules which are: Data Collection, Normalization, Orchestration, Network Integration and Service Discovery. So WAE manages network traffics by taking request type, network saturation, CPU usage and existing containers in consideration. 
\item We collaborate with CDN company (Medianova) to evaluate results in live environment. Our comparison results can be used in future works to deploy complex virtualized CDN systems.
\end{itemize}
The rest of this paper is organized as follows. The proposed architecture and CDN-specialized container orchestration are explained in sec. II. The implementation and performance analysis are shown in sec. III and sec. IV respectively. We conclude the paper in sec. V.

\section{The Container-based Content Delivery Network with Resource Provisioning}
Following subsections firstly describe the general view of our design by comparison with traditional architecture and then illustrates the implementation of our system and and the Workload Automation Engine. 

\subsection{Architecture}
\begin{figure}
    \centering
    \subfigure[before: Traditional Content Delivery Network architecture]
    {
        \includegraphics[width=1.58in]{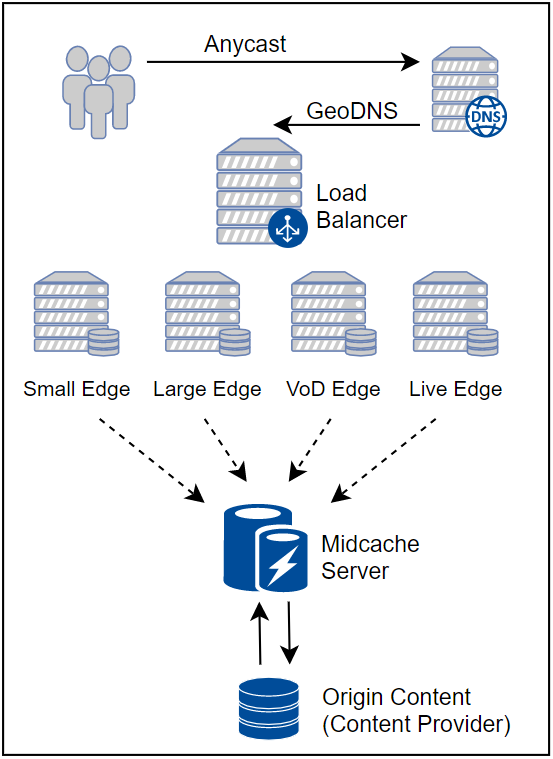}
        \label{fig:before}
    }
    \subfigure[after: Container-based virtualized Content Delivery Network]
    {
        \includegraphics[width=1.58in]{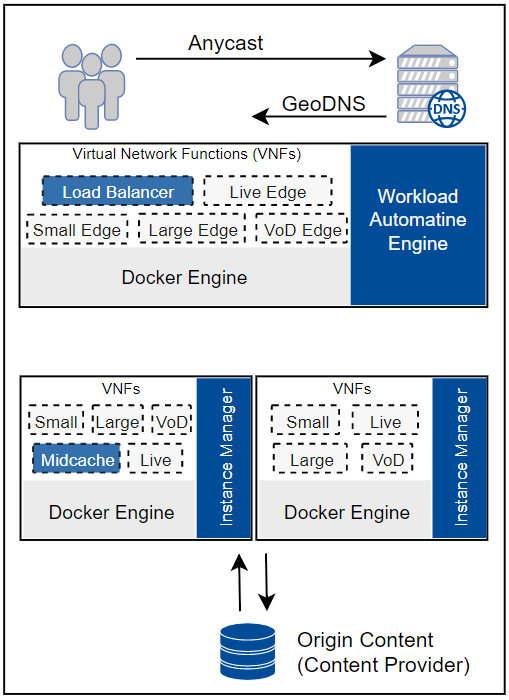}
        \label{fig:after}
    }
    \caption{Architecture Comparison}
    \label{fig:comparison}
\end{figure}

The sample traditional PoP which we work on contains one DNS server, one load balancer, four distinct edge servers and one mid-cache with an origin server as seen in Fig.  \ref{fig:before}. CDN edge servers also known as cache servers provide different service in terms of request type and we group the edge servers in four categories which are: (i) \textit{Small Edge}: serves small files ($\lessapprox$1MB, e.g., picture), (ii) \textit{Large Edge}: serves large data ($\gtrapprox$1MB, e.g., pdf document), (iii) \textit{VoD Edge}: responsible for VoD streaming and (iv) \textit{Live Edge}: responsible for live streaming. This grouping is also preserved in the containerized architecture. 

Our virtualized network architecture (see Fig. \ref{fig:after}) includes all CDN roles in the form of VNFs on the underlying Docker containerization. Also it contains WAE and instance managers (IM) to orchestrate a part of VNFs with communicating over interfaces. Thus our proposed model overcome intense requests with flexible manner also with less physical machine in total. For example, Fig. \ref{fig:before} contains only one edge server for small requests; however, in our design in Fig. \ref{fig:after} client requests can be meet with up to 3 virtual edge server in heavy load. Also our design focus on the orchestration of only edge servers which are directly related to QoS. Detailed system design and the algorithm is explained in below subsections. 
\subsection{System Model}
\subsubsection{Instance Manager (IM)} 
\begin{figure}[t!]
  \centering
  \includegraphics[width=\linewidth]{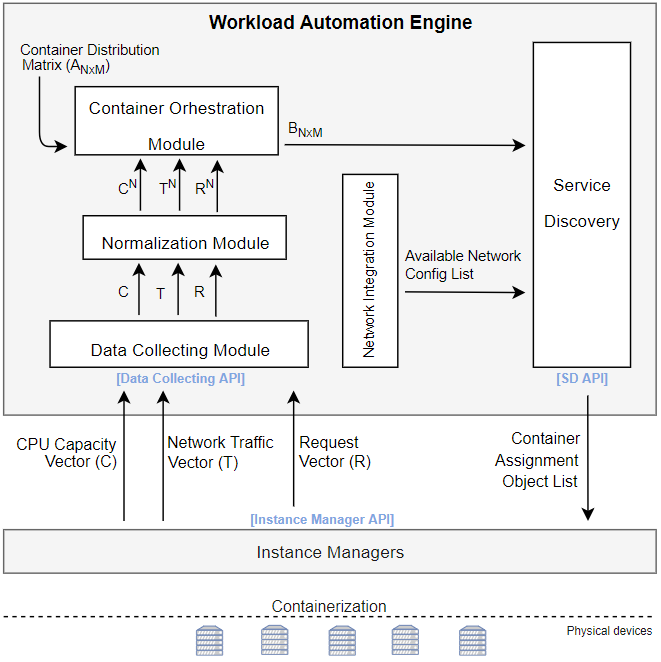}
  \caption{System Model}
  \label{fig:systemModel}
\end{figure}

Each machine in the virtualized PoPs includes Instance Manager which provides southbound interface through Data Collection API (DC API). Each IM in the PoP sends CPU capacity and output network traffic of the host machine to WAE. Also IMs supply a vector for number of request coming to edge servers. Here, number of request means that four edge servers collect the coming request within predefined period (for example ten minutes) then IMs get the total summation from each different edge servers to produce a vector for total number of request with respect to request type (small requests, large requests, streaming and live requests). Fig. \ref{fig:systemModel} shows the interfaces and supplied inputs by IM.
\subsubsection{Workload Automation Engine (WAE)} 
We implement WAE (Fig. \ref{fig:systemModel}) which manages containers with respect to CPU usages of host machines, outgoing network traffic, request URL types and existing container; then produces an output to send new VNF assignments to Instance Manager APIs. Thereby, CDN base specialized dynamic scaling improves existing virtualization solutions by re-arranging PoP according to client demand.

Workload automation engine contains five modules which are: Data Collection, Normalization, Orchestration, Network Integration and Service Discovery. The system design of five modules are explained in the following, respectively.

\paragraph{Data Collection Module} Data Collection RESTful API listen IM interfaces, which is placed in different machines, and aggregates input data for resource provisioning algorithm. Output of DC Module includes three vector which are \textbf{C}, \textbf{T} and \textbf{R}; CPU capacity vector, output link network traffic, number of requests with request types, respectively. 
\paragraph{Normalization Module} Output of the DC Module will be input for Normalization Module. The vectors \textbf{C} and \textbf{T} are rescaled to range of \(\mathnormal{[0,1]}\) for transforming data from unit based to percentage base form. Likewise, the vector \textbf{R} is an input for function $ f(x_i)=\frac{x_i} {\sum\limits_{j=1}^{i} x_i} $ to obtain weighted distribution of requests. As a result, we obtain pure input for resource provisioning algorithm.

\paragraph{Container Orchestration Module}
The module includes resource provisioning algorithm to orchestrate containers on client demand. The basic heuristic behind this algorithm is to dynamically scale load for heterogeneous client requests in consideration of existing resources. Meeting client requests with minimized set of containers (1) lead to formulating the following optimization. Constraint (2) ensures that resource distribution is kept inside threshold ($\pm$ 0.1), and (3) guarantees that each physical machine has at most 1 virtual function for each type. We indicate the mathematical notations of network parameters and system arguments in Table \ref{tb_math}.  

\begin{table}[ht]
\renewcommand{\arraystretch}{1.3}
\caption{Mathematical Notations}
\label{tb_math}
\centering
\begin{tabular}{l|l}
\hline
\bfseries Parameter & \bfseries Definition\\
\hline
N & number of physical machines\\
M & \begin{tabular}{@{}l@{}}number of different network functions to \\ orchestrate\end{tabular}  \\
Z & total number of containers\\
A\textsubscript{NxM} & \begin{tabular}{@{}l@{}}2-dimensional container distribution matrix \\ (input)\end{tabular} \\
B\textsubscript{NxM} & \begin{tabular}{@{}l@{}}2-dimensional container distribution matrix \\ (output)\end{tabular} \\
R = \{r\textsubscript{1}, r\textsubscript{2}, r\textsubscript{3}, r\textsubscript{4}\} & request vector\\
C = \{c\textsubscript{1}, c\textsubscript{2}, \ldots, c\textsubscript{N}\} & cpu capacity vector\\
T = \{t\textsubscript{1}, t\textsubscript{2}, \ldots, t\textsubscript{N}\} & output link network traffic vector\\
\hline
\end{tabular}
\end{table}

\begin{align}
& \underset{Z}{\text{minimize}}
& & \sum\limits_{i\in N} Z\textsubscript{i}  \label{eq:optimization}\\
& \text{subject to}
& & X_i^N - 0.1 \leq R_i^N, \leq X_i^N + 0.1, \quad \forall i \in \mathbb{N} \\
& & & A_{ij} = 1 \quad \vee \quad A_{ij} = 0, \quad \forall i,j \in \mathbb{N}
\end{align}

where X is summation of vectors \textbf{C} and \textbf{T} multiplied by matrix \textbf{A}, that is \(\mathnormal{X = (C+T)^\intercal \cdot A}\). On the other hand, superscript \(\mathnormal{N}\) represents the normalized vector mentioned in Normalization Module.

  \begin{algorithm}
    \algsetup{linenosize=\tiny}
  \scriptsize
 \caption{CDN-specialized Container Orchestration}\label{algo}
 \begin{algorithmic}[1]
 \renewcommand{\algorithmicrequire}{\textbf{Input:}}
 \renewcommand{\algorithmicensure}{\textbf{Output:}}
 \REQUIRE average CPU capacity vector \( \mathnormal{C} \), network output traffic \( \mathnormal{T} \), number of request distribution for each request type \( \mathnormal{R^N}\), and container distribution matrix \( \mathnormal{A_{NxM}} \)
 \ENSURE  new container distribution matrix \( \mathnormal{B_{NxM}} \)
 \\ \STATE{\(\mathnormal{V_{sum}} \gets \mathnormal{C} + \mathnormal{T} \)  }
 \\ \STATE{\(\mathnormal{D} \gets \mathnormal{V_{sum}} \cdot \mathnormal{A} \)  }
  \\ \STATE{\(\mathnormal{D^N} \gets\sfunction{Normalize}(D)\)  }
\WHILE{\( \mathnormal{A_{NxM}} \) is changed}
\FOR{$i = 1$ to $length$ $of$ $D^N$}
  \IF{$R_i^N$ - 0,1 \textgreater $D_i^N$}
 	\STATE $j\gets\sfunction{FindMinLoadedMachine}(\mathnormal{A_{NxM}}, \mathnormal{V_{sum}})$
    \STATE $A_{ji}\gets1$
  \ELSIF{$R_i^N$ + 0,1 \textless $D_i^N$}
 	\STATE $j\gets\sfunction{FindMaxLoadedMachine}(\mathnormal{A_{NxM}}, \mathnormal{V_{sum}})$
    \STATE $A_{ji}\gets0$
  \ENDIF
\ENDFOR
 \\ \STATE{\(\mathnormal{D} \gets \mathnormal{V_{sum}} \cdot \mathnormal{A} \)  }
  \\ \STATE{\(\mathnormal{D^N} \gets\sfunction{Normalize}(D)\)  }
\ENDWHILE

 \end{algorithmic} 
 \end{algorithm}

Algorithm \ref{algo} shows the algorithmic representation of the optimization problem (\ref{eq:optimization}) to find minimum number of containers in the system to serve client request under the resource capacity: CPU and Network Link. In this perspective, our algorithm operates as follow: two resource capacities, CPU and network output traffic, are equally important for this case so that both of them are sum up (line 1) to obtain total resource capacity. Then, the algorithm calculates resource distribution over containers (line 2) to get resource distribution over virtual edge servers. Moreover, remaining part of algorithm try to find which type of virtual edge server require expansion or which ones should be shrinking. For this, each density distribution of resources \(D^N\) is compared with the density distribution of client requests \(R^N\) with respect to threshold \(0.1\) (line 6 and 9). According to condition, the algorithm decide a physical machine to create or drop the container with respect to resource usage (minimum loaded or maximum loaded) (line 7 and 10). The algorithm calculates new density distribution of resources \(D^N\) with recently calculated \(A\) at the end of each iteration also line 4 guarantees that the algorithm is terminated.  
 
\paragraph{Network Integration Module}
In our design, we want to configure VNFs with public IPs so that it can be reachable remotely. Also it will be well-suited for traditional CDN routing principles. Network Integration Module provide available network configurations list for new deployed containers and manages network configurations of dropped containers. Network Integration Module get dropped containers list from Service Discovery Module and send available IP list to Service Discovery Module.    

\paragraph{Service Discovery Module (SD API)}
Each machine has all type of containers with status of paused or running which is controlled by Service Discovery Module. Service Discovery is northbound interface to integrate underlying physical infrastructure with virtual network functions. SD API supply list of container assignment object according to decision of orchestration algorithm and network configurations provided by Network Integration Module. Container assignment objects for each machine is send to IMs. 

\section{Implementation}
As a proof of concept, we deploy three host machine with 8-core Intel\textsuperscript{\textcopyright} processor, 16 GB of RAM, and running CentOS\textsuperscript{\textcopyright} 7.4. Each host machines are configured to include VNF instances as Docker container with the version of 17.09. We are using \textit{nginx} to serve HTTP connections in the virtual edge servers, with version of \textit{nginx} 1.12. As we mentioned in our system design, VNF instances are accessible publicly so that we use \textit{macvlan} network driver to use CDN edge servers as capable of direct routing. The \textit{macvlan} configuration we used is shown below.

\begin{lstlisting}[frame=single,basicstyle=\ttfamily]
docker network create -d macvlan --subnet=
<desired subnet> --gateway=<gateway> -o 
parent=eth0 mn-container-network
\end{lstlisting}

\begin{figure*}[ht]
  \centering
  \includegraphics[width=0.9\linewidth]{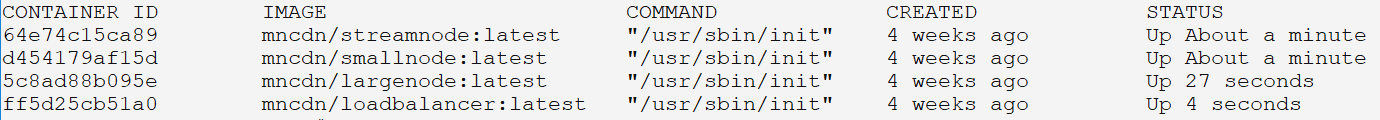}
  \caption{System Screen from Proposed Network Architecture}
   \label{fig:containerizedArch}
\end{figure*}

Moreover, we create \textit{Dockerfile} for CDN roles which means we write discretely five \textit{Dockerfile} on the private Docker Hub. Also Algorithm \ref{algo} is implemented and run in every 10 minutes periods to re-orchestrate containers. We implement containerized CDN network as we described in \ref{fig:after}. Each machines include VNFs of CDN roles, for example Fig. \ref{fig:containerizedArch} shows \textit{Docker} instances of the master machine.    

\section{Performance Evaluation}
We compare our proposed design with traditional CDN PoP, which includes bare-metal servers for each CDN functionalities, to demonstrate performance results. Moreover, we indicate the importance of orchestration by comparing with the virtualized topology which is not include our orchestration module. 

\begin{figure}[ht]
  \centering
  \includegraphics[width=\linewidth]{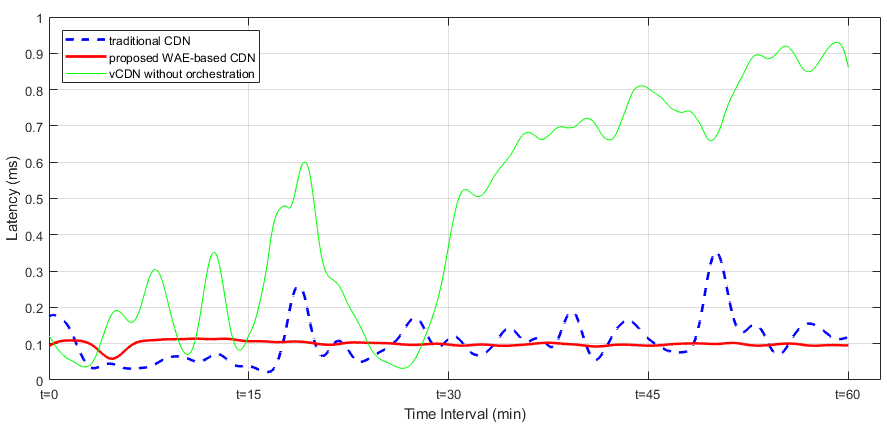}
  \caption{Latency Comparison}
   \label{fig:latency}
\end{figure}

We choose 1-hour client requests from the observations on the topology in Vienna. Then, we generate the same requests with \textit{Jmeter}, one of the open source performance test tool written by Java. As a result, our scenario contains:
\begin{itemize}
\item \textit{Image Request} 5000 users in total with 5 second ramp up time, which means that it start with 1000 users and it reach 5000 users in 5 seconds.
\item  \textit{Large Content Request} 1500 users in total with 3 second ramp up time
\item \textit{Stream Request} 500 users in total with 5 second ramp up time
\end{itemize} 
We observe that the latency of the proposed heuristic is roughly 45\% less than traditional bare-metal CDNs. In addition, as seen in Fig. \ref{fig:latency}, the proposed model makes difference with orchestration module compared with containerized CDN which is not include resource provisioning heuristic. 

\begin{table}[ht]
\centering
\caption{Average Resource Usage in Traditional CDN}
\label{table:tra-resource}
\begin{tabular}{|l|l|l|}
\hline
\textbf{CDN Role} & \textbf{CPU Usage} & \textbf{\begin{tabular}[c]{@{}l@{}}Network Output \\ Link Usage\end{tabular}}  \\ \hline
small edge server & 7\% & 0.2\%  \\
large edge server & 5\% & 12\% \\
stream edge server & 2.5\% & 3\% \\
load balancer & 1\% & 2\% \\
\hline
\end{tabular}
\end{table}

\begin{table}[ht]
\centering
\caption{Average Resource Usage in Container-based CDN}
\label{table:conta-resource}
\begin{tabular}{|l|l|l|}
\hline
\textbf{Host Machines} & \textbf{CPU Usage} & \textbf{\begin{tabular}[c]{@{}l@{}}Network Output \\ Link Usage\end{tabular}}  \\ \hline
physical machine 1  & 32\% & 48\%  \\
physical machine 2 & 28\% & 42\% \\
physical machine 3 & 25\% & 51\% \\
\hline
\end{tabular}
\end{table}

\begin{figure}[!ht]
    \centering
    \subfigure[Cost Comparison]
    {
        \includegraphics[width=0.46\linewidth,height=3cm]{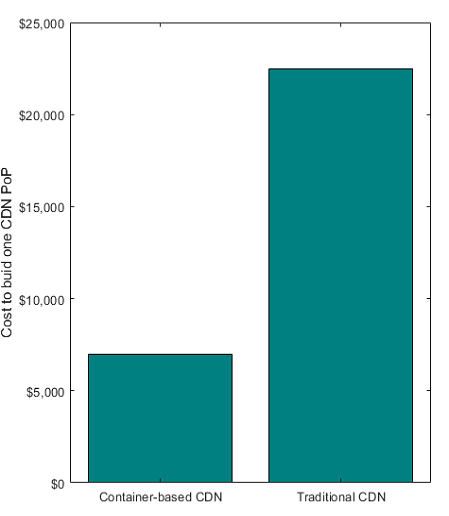}
   		\label{fig:cost}
    }
    \subfigure[Resource Usage Comparison]
    {
        \includegraphics[width=0.47\linewidth,height=3cm]{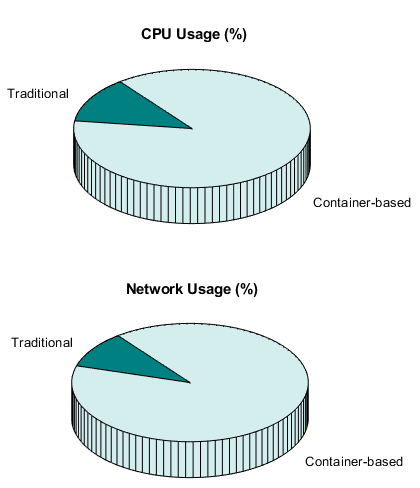}
   		\label{fig:piechart_resource}
    }
    \caption{Cost and Resource Usage Comparison}
    \label{fig:cost_and_resource}
\end{figure}

As seen in Table \ref{table:tra-resource} and Table \ref{table:conta-resource} , we obtain approximately 20\% more CPU efficiency and 35\% more utilized network in proposed model. In other words, we reach less latency with less resource capacity thanks to resilient resource provisioning heuristic. On the other hand, as seen in Table \ref{table:conta-resource}, the resources of our proposed CDN architecture does not saturate so it can handle with more heavy load. Graphical representation is placed in Fig. \ref{fig:piechart_resource}.

Moreover, we reduce the deployment cost of the PoP by three times compared to traditional architecture as seen in Fig. \ref{fig:cost}

As seen in the results, WAE enhance existing containerized CDN using orchestration module and produce steady and decreased latency. Because WAE uses resources and existing containers as inputs, also takes client trends in consideration (not in other orchestration solutions). Thereby, CDN-specialized orchestration uses less resources with more effective usage. This leads to three times less service deployment.     

\section{Conclusion}
In this paper, we proposed a resilient resource provisioning heuristic for virtualized CDN to provide steady service for end-users. In our approach, workload automation engine intelligently performs VNF assignments to achieve minimum latency with least deployment cost. Also, we collaborated with Medianova to observe client requests in live environment, then to test our proposed design. According to results, our design reduces the latency by 45\% with 66\% percentage decreased deployment cost compared to traditional CDNs. Also we optimized the CPU and Network up to 20\% and 35\% percentage, respectively. In future work, we concentrate on orchestration framework for large scale CDN topologies to deploy dynamically changeable CDN PoPs, which are interconnected each other.

\end{document}